\documentstyle[aps,prl,floats,psfig,amssymb]{revtex}
\catcode`@=11
\def\references{%
\ifpreprintsty
\bigskip\bigskip
\hbox to\hsize{\hss\large \refname\hss}%
\else
\vskip 24pt
\hrule width\hsize\relax
\fi
\list{\@biblabel{\arabic{enumiv}}}%
{\labelwidth\WidestRefLabelThusFar  \labelsep4pt %
\leftmargin\labelwidth %
\advance\leftmargin\labelsep %
\ifdim\baselinestretch pt>1 pt %
\parsep  4pt\relax %
\else %
\parsep  0pt\relax %
\fi
\itemsep\parsep %
\usecounter{enumiv}%
\let\p@enumiv\@empty
\def\theenumiv{\arabic{enumiv}}%
}%
\let\newblock\relax %
\sloppy\clubpenalty4000\widowpenalty4000
\sfcode`\.=1000\relax
\ifpreprintsty\else\small\fi
}
\catcode`@=12
\draft

\def\lsim{\mathrel{\raise.3ex\hbox{$<$\kern-.75em\lower1ex\hbox{$\sim$}}}}
\def\gsim{\mathrel{\raise.3ex\hbox{$>$\kern-.75em\lower1ex\hbox{$\sim$}}}}

\begin{document}

\twocolumn[\hsize\textwidth\columnwidth\hsize\csname
@twocolumnfalse\endcsname

\hfill\vbox{
\hbox{MADPH-01-1241}
\hbox{hep-ph/0108261}
\hbox{}}

\title{Testing radiative neutrino mass generation via
$R$-parity violation at the Tevatron}
\author{V.~Barger, T.~Han, S.~Hesselbach and D.~Marfatia}
\address{Department of Physics, University of Wisconsin, Madison, WI 53706, 
USA}

\maketitle

\begin{abstract}
An $R$-parity violating SUSY model with lepton-number violating couplings
$\lambda'_{i33}$, $i=2,3$ can generate a neutrino
mass spectrum that explains the recent results from neutrino
oscillation experiments.
These $R$-parity violating couplings lead to a clean signal with
 at least one isolated lepton and at least three tagged $b$ jets that
is accessible in
chargino and neutralino production 
at the Tevatron collider. This signature 
can be probed at $3 \sigma$ up to
$m_{1/2} = 230$~GeV (320~GeV) with an integrated luminosity of 
$ 2~\mathrm{fb}^{-1}$ ($ 30~\mathrm{fb}^{-1}$).

\end{abstract}]

\narrowtext

There is now strong evidence from both atmospheric~\cite{atm} and 
solar~\cite{solar,sno} neutrino experiments that neutrinos are not massless. 
The $\nu_\mu$ flux deficit and zenith angle dependence 
seen in atmospheric experiments is well-explained if 
$\nu_\mu \rightarrow \nu_\tau$ oscillations occur with maximal mixing and
$\delta m^2 \sim 3.5 \times 10^{-3}$ eV$^2$. The recent results from the
SNO experiment~\cite{sno}, 
give compelling confirmation that $\nu_e$ neutrinos from the
Sun oscillate to $\nu_{\mu,\tau}$ neutrinos with large mixing and quite 
possibly $\delta m^2 \sim 5 \times 10^{-5}$ eV$^2$~\cite{snores}. 
Thus, it is important
to find a testable explanation of these mixing angles and 
mass-squared differences. The most promising extension of the Standard Model
is obtained by its supersymmetrization. In the
Minimal Supersymmetric Standard Model (MSSM) a discrete symmetry called
$R$-parity is usually assumed to be conserved.
However, if this discrete symmetry is broken, 
radiative neutrino mass generation is a direct consequence
of the MSSM~\cite{hall}. 
In this Letter, we show that this alternative
to the see-saw mechanism~\cite{see-saw} can be subject to direct tests
at the Tevatron collider.

\noindent
\underline{Radiative neutrino mass generation}

The most general superpotential allowed by SM gauge symmetry and
supersymmetry contains the explicit $R$-parity violating terms
\begin{equation}
W_{\not\!\,R} = \lambda_{ijk} L_i L_j \bar{E}_k +
    \lambda'_{ijk} L_i Q_j \bar{D}_k +
    \lambda''_{ijk} \bar{U}_i \bar{D}_j \bar{D}_k \,,
\end{equation}
where $i,j,k = 1,2,3$ are generation indices and $L$, $Q$, $E$, $D$
and $U$ denote the left-handed lepton and quark doublets and the
right-handed charged-lepton down- and up-type quark singlets,
respectively \cite{bargerdreiner}. 
The first two terms ($\lambda$, $\lambda'$) violate lepton-number and the
third term ($\lambda''$) baryon-number conservation.

\begin{figure}[t]
\psfig{file=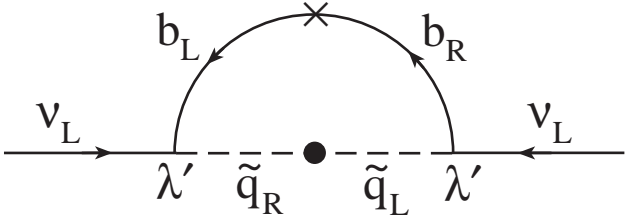}
\caption{\label{feyn} The dominant one loop diagram via which Majorana
mass terms are generated for left-handed neutrinos. }
\end{figure}

An important consequence of the $\lambda'$ interactions is the generation
of Majorana masses for left-handed neutrinos through quark-squark loops as in 
Fig.~\ref{feyn}. Similar diagrams result from the $\lambda$ couplings. 
The dominant contribution to neutrino masses comes from $b\tilde{b}$ 
loops, so we focus on the $R$-parity breaking interactions 
resulting from $\lambda'_{i33}$: 
\begin{eqnarray}
{\cal L}/\lambda'_{i33}  &=& 
 \tilde{\nu}_L^i \bar{b}_R b_L + \tilde{b}_L \bar{b}_R \nu_L^i
  + (\tilde{b}_R)^* (\bar{\nu}_L^i)^c b_L
  - \tilde{e}_L^i \bar{b}_R t_L \nonumber\\ &&
 - \tilde{t}_L \bar{b}_R e_L^i
  - (\tilde{b}_R)^* (\bar{e}_L^i)^c t_L  + \textrm{h.c.}
\end{eqnarray}

Specific models involving $\lambda'_{i33}$ 
that accommodate maximal $\nu_\mu-\nu_\tau$ 
mixing in atmospheric neutrino oscillations and the Large Mixing Angle 
solution to the solar neutrino anomaly have been constructed in 
Ref.~\cite{drees}. The mass matrix arising from these models takes the 
form
\begin{equation}
m_{\nu} \sim {3\over 8 \pi^2} {\lambda'}^2 {m_b^2 \over m_{\tilde q}}\left(\begin{array}{ccc}
\epsilon^2 & \epsilon &\epsilon \\
\epsilon & 1 & 1 \\
\epsilon & 1 & 1
\end{array}\right)
\label{matrix}
\end{equation}
where $\epsilon=m_s/m_b$.  
The explicit breaking of a discrete symmetry that automatically
forbids baryon number violating interactions and bilinear
$R$-parity violating couplings leads to a suppression of the
$\lambda'_{133}$ coupling by $\epsilon$ relative
to $\lambda'_{233}$ and $\lambda'_{333}$.
Although the bilinear terms are generated at one-loop,
their contributions to the neutrino mass matrix of Eq.~(\ref{matrix})
are estimated to be small, suppressed by a loop-factor squared.
With
\begin{equation} 
\lambda'_{233} \simeq \lambda'_{333} \sim 10^{-4} 
\bigg({m_{\tilde q} \over 200\, {\rm {GeV}}}\bigg)^{1/2}\,,
\end{equation}
the largest mass is ${m_{\nu_3}} \sim 0.06$ eV,
corresponding to the atmospheric scale $\sqrt{\delta m^2_{{\rm atm}}}
$~\cite{atm}.
The values of these couplings are well within the existing $2\sigma$ limits: 
$\lambda'_{233}\,,  \lambda'_{333}$ of ${\cal O} (0.1)$ and
$\lambda'_{133}$  of ${\cal O} (10^{-3})$~\cite{tevatron}. The two other
eigenvalues of the matrix in Eq.~(\ref{matrix}) are shifted from zero
by corrections of ${\cal O} (\epsilon)$, thereby yielding the correct
ratio for $\delta m^2_{{\rm solar}}/\delta m^2_{{\rm atm}} \sim \epsilon^2$.
  
\noindent
\underline{Production and decay of the LSP}

Because the $R$-parity violating Yukawa couplings are small,
we assume the sparticles have the same mass patterns as
in the mSUGRA model~\cite{run2SUSY} and the lightest 
neutralino ($\tilde{\chi}^0_1$) is the 
lightest supersymmetric particle (LSP). 
The production and decay of the sparticles occur dominantly
through the $R$-parity conserving channels except for the decays of the
LSP. 

The dominant SUSY production processes for heavy squarks and gluinos
at the Fermilab Tevatron are
pair production of the lightest chargino ($q \bar{q} \to
\tilde{\chi}^+_1 \tilde{\chi}^-_1$) and associated production of the
lightest chargino with the two lightest neutralinos ($q \bar{q}' \to
\tilde{\chi}^\pm_1 \tilde{\chi}^0_i\,,\, i=1,2$) \cite{run2SUSY}.
The $\tilde{\chi}^\pm_1$ and $\tilde{\chi}^0_2$ decay in their
$R$-parity conserving channels into the LSP. 

\begin{figure}[t]
\centering
\mbox{\psfig{file=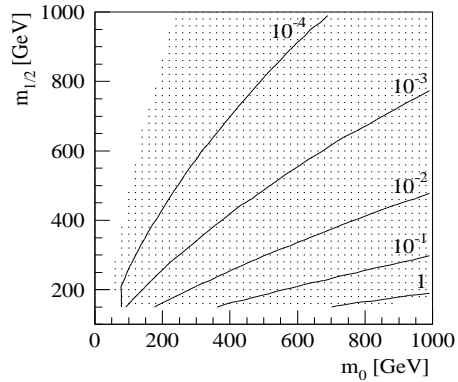,width=6cm,height=5.25cm}}
\caption{\label{decaylen}Contours of the rest decay length $c\tau$ (in meters)
of the LSP in the $m_0$-$m_{1/2}$ parameter space for 
$\lambda'_{233} = \lambda'_{333} = 10^{-4}$.
The other parameters are fixed at
$A_0=0$, $\tan\beta=10$ and $\mu>0$. The dots represent mSUGRA points
that satisfy $m_{\tilde{\chi}^\pm_1}\,,m_{\tilde{e}_R}>100$~GeV,
 $m_{\tilde{\tau}_1}>85$~GeV and $m_h>90$~GeV.}
\end{figure}

Figure~\ref{decaylen} shows contours of the rest decay length $c\tau$ 
(in meters) of
the LSP in the $m_0$-$m_{1/2}$ parameter space for 
$\lambda'_{233} = \lambda'_{333} = 10^{-4}$.
Since $\lambda'_{133}$ is two orders of
magnitude smaller than $\lambda'_{233}$ and $\lambda'_{333}$, 
its effect on the branching ratio of the LSP can be safely ignored.
Except in scenarios with $m_0 \gg m_{1/2}$, 
the LSP decays well inside the detector.
In general the decay length scales according to $c\tau \sim
(\lambda'_{i33})^{-2}$.
For $m_{\tilde{\chi}^0_1} < m_t + m_b$, the LSP decays to
${\tilde{\nu}_i}^* \nu_i$ and ${\tilde{b}}^* b$ lead to 
a neutrino and a $b \bar{b}$ pair:
\begin{equation}
\tilde{\chi}^0_1 \to \nu_\mu b \bar{b}, \nu_\tau b \bar{b} +
 \textrm{c.c.}
\end{equation}
For $m_{\tilde{\chi}^0_1} > m_t + m_b$, the decay channels involving a
lepton and a top quark are also open
($\tilde{\chi}^0_1 \to 
{\nu_\mu b \bar{b}, \nu_\tau b \bar{b},
 \mu t \bar{b}, \tau t \bar{b}}
 + \textrm{c.c.}$). For $m_{1/2} \lesssim 500$~GeV the top channels have no appreciable
branching fraction and even for $m_{1/2} < 1000$~GeV their branching fractions
are always smaller than $\sim 20\%$.
Hence we focus on the decay of the LSP
into a neutrino and a $b \bar{b}$ pair.
Since  the sparticles are produced in pairs
and decay into the LSP through the usual $R$-parity conserving
channels, there are two $\tilde{\chi}^0_1$ LSPs in each
supersymmetric event. Thus we have the promising signature of four $b$
jets and missing energy.

\noindent
\underline{Signals and SM backgrounds}

We simulate SUSY events at the parton level
with the event generator PYTHIA, version
6.157, which includes the supersymmetric extension SPYTHIA
\cite{pythia} and a further extension which provides the $R$-parity violating
decays of the LSP \cite{rpythia}. 
The masses of the SUSY particles are computed from the mSUGRA input
parameters $m_0$, $m_{1/2}$, $\tan\beta$, $A_0$ and $\rm sign (\mu)$
using the two loop RGEs of the program ISAJET \cite{isajet} and then
used as electroweak scale input parameters for PYTHIA.

The detector energy resolutions at the Tevatron are simulated with
\cite{runii}
\begin{eqnarray}
\Delta E_j/E_j &=& 0.8/\sqrt E_j \quad {\rm for\ hadrons}\,,\nonumber\\ 
\Delta E_e/E_e &=& 0.2/\sqrt E_e \quad {\rm for\ electrons}\,,
\end{eqnarray}
and the smearing of muons is neglected. 
We add the smeared transverse momenta of all the particles to obtain the
missing transverse momentum $\overlay{/}{p}_T$ of the event.
Then we apply the ``basic cuts''
\begin{eqnarray}
p_T >15~\mbox{GeV},\ \ |\eta|<2\,,\ \ \Delta R>0.5
\label{basic}
\end{eqnarray}
on the $b$ and $c$ quarks in the event and 
simulate the $b$-tagging efficiency by accepting the surviving $b$
quarks with $50\%$ probability and the misidentification of
$c$ by accepting the surviving $c$ quarks with $5\%$
probability as $b$ jets.
We ensure the isolation of the $b$ jets by counting two $b$ with
$\Delta R_{bb}<0.5$ as one $b$ jet.
To take advantage of the large number of $b$ quarks in the SUSY events
and to suppress the background, we consider two signatures with at least
three tagged $b$ jets and
\begin{enumerate}
\addtolength{\itemsep}{-2.5mm}
\item{at least one isolated lepton ($l=e,\mu$) which survives the basic cuts
$p_T >10~\mbox{GeV},\ |\eta|<2$
\mbox{($\geq 3b,\, \geq 1l$ signature) or}
\item{
a missing transverse momentum $\overlay{/}{p}_T > 20$~GeV
(\mbox{$\geq 3b,\,\overlay{/}{p}_T$ signature).}} 
}
\end{enumerate}
The lepton isolation is ensured by counting only leptons with no other
particle in a cone $\Delta R<0.5$ around the lepton.
The $\geq 3b,\,\overlay{/}{p}_T$ signature has the advantage that all
the supersymmetric production processes contribute regardless of the
 particles produced and their decays.

The major background process is $t \bar{t}$ production where one of
the $W$ decays via $W^- \to b \bar{q},\ (q=c,u)$ or a $c$ quark from $W^-
\to s \bar{c}$ is misidentified. The branching ratio of the former decay
is very small so the background from the latter decay
and misidentified $c$ dominates.  There is, in addition, a potentially 
dangerous background for the $\geq 3b,\,\overlay{/}{p}_T$ 
signal from the QCD production of $b\bar b b\bar b,\ b\bar b c\bar c$.
With the basic cuts of Eq.(\ref{basic}), the background rate from this 
process with three
tagged $b$ jets is about 5 pb, which is more than an order of magnitude
higher than our typical signal rate. Although there is no significant
$\overlay{/}{p}_T$ from the process, the imperfect jet energy measurement
and $b$ semileptonic decay  can result in some $\overlay{/}{p}_T$. However, 
we expect
that our cut $\overlay{/}{p}_T>20$ GeV and other kinematical cuts 
would significantly suppress this reducible QCD background. More detailed 
simulations of this background to the $\geq 3b,\,\overlay{/}{p}_T$ 
signal need to be implemented. 

There are kinematical differences between the signal
and the $t \bar{t}$ background that can be exploited to separate them. 
In particular, the distributions of the number of $b$ jets and 
 transverse momenta are different for the signal and background.
 For purposes of illustration, we choose an mSUGRA reference scenario
\begin{eqnarray}
m_{1/2}&=&m_0=200~\textrm{GeV}\,,\ 
A_0=0\,,\ 
\tan\beta=10\,,\ 
\mu>0\,, \label{refszen}
\end{eqnarray}
for which the masses of the neutralinos and charginos are
\begin{eqnarray}
m_{\tilde{\chi}^0_1} = 75~\textrm{GeV}\,,\ 
m_{\tilde{\chi}^0_2} = 135~\textrm{GeV}\,, \ 
m_{\tilde{\chi}^+_1} = 133~\textrm{GeV}\,.
\end{eqnarray}

\begin{figure}[t]
\mbox{
\psfig{file=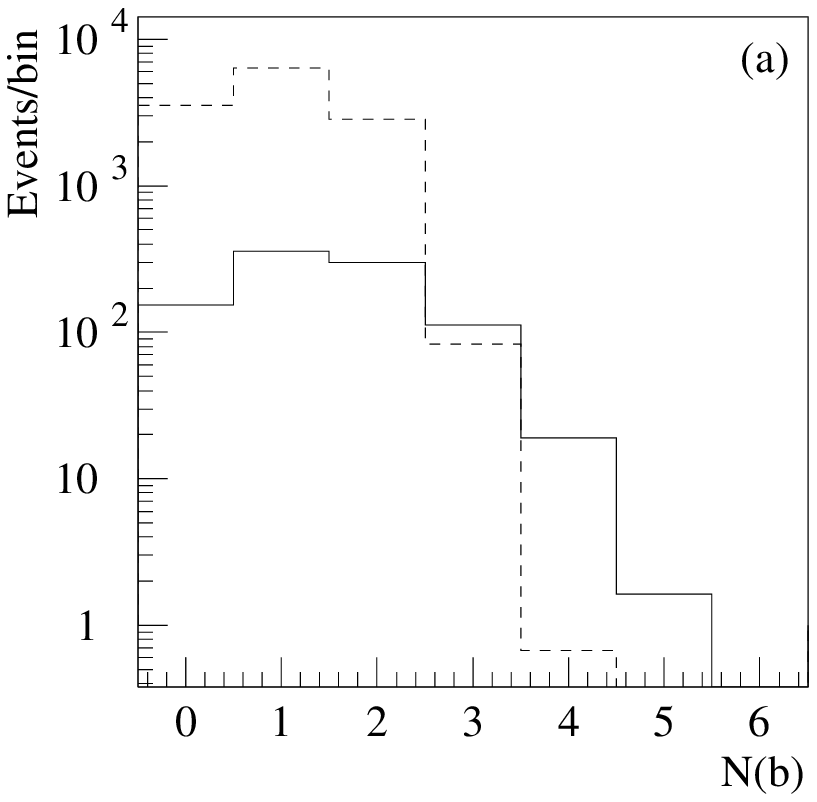,width=4.2cm}
\psfig{file=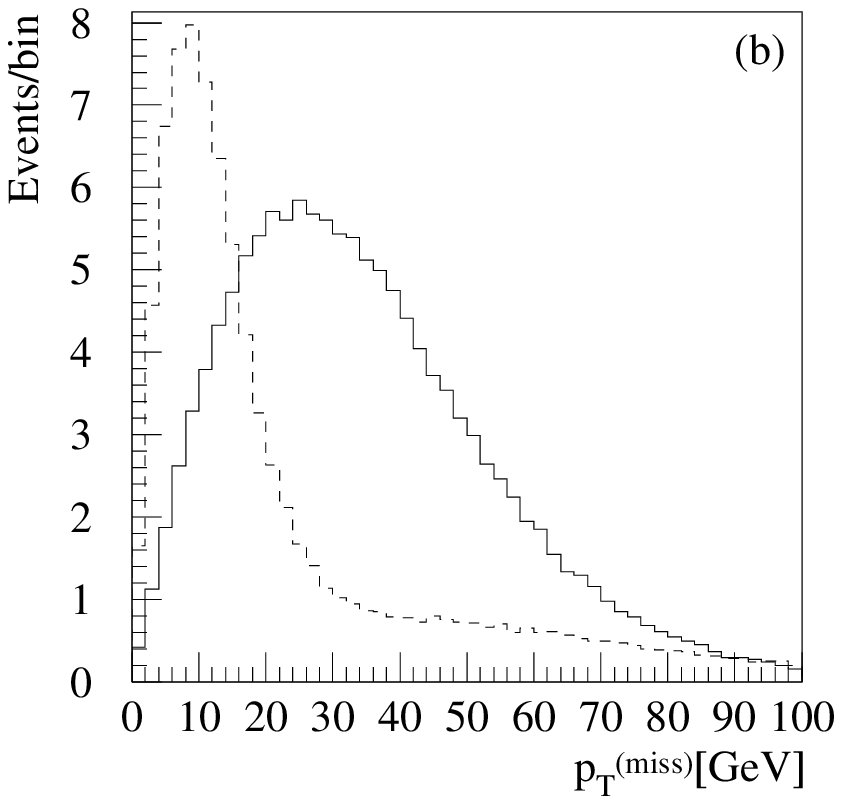,width=4.2cm}
}
\caption{\label{nbdist}Distributions (a) of the number of tagged $b$ jets
and (b) of the missing transverse energy
$\overlay{/}{p}_T$ in events with at least three tagged $b$ jets
for the SUSY production processes
in the reference scenario of Eq.~(\ref{refszen}) (solid) and
for the $t \bar{t}$ background (dashed). The distributions are
 normalized to the expected
number of events at Run II of the Tevatron with $ 2~\mathrm{fb}^{-1}$.}
\end{figure}

Figure~\ref{nbdist}(a) shows the distribution of the number of tagged $b$
jets from the $R$-parity violating decay of the LSP in the reference
scenario of Eq.~(\ref{refszen}) and from the $t \bar{t}$ background. Here and
henceforth, we represent the signal with a solid line and the background
with a dashed line. 
While the background for events with 0, 1 or 2 tagged $b$ jets is more
than 10 times as large as the SUSY signal, the requirement of at least
three tagged $b$ jets yields a signal-to-background event ratio $S/B =
133/84$ for the expected luminosity $ 2~\mathrm{fb}^{-1}$ of
the Tevatron Run II even without further cuts.

Figure~\ref{nbdist}(b) gives the distributions
of the missing energy $\overlay{/}{p}_T$ in events with at least three tagged
$b$ jets.
The background is peaked for low missing energy, hence the cut
$\overlay{/}{p}_T > 20$~GeV significantly suppresses the background in
the $\geq 3b,\,\overlay{/}{p}_T$ signature, yielding $S/B = 100/31$.

\begin{figure}[t]
\mbox{
\psfig{file=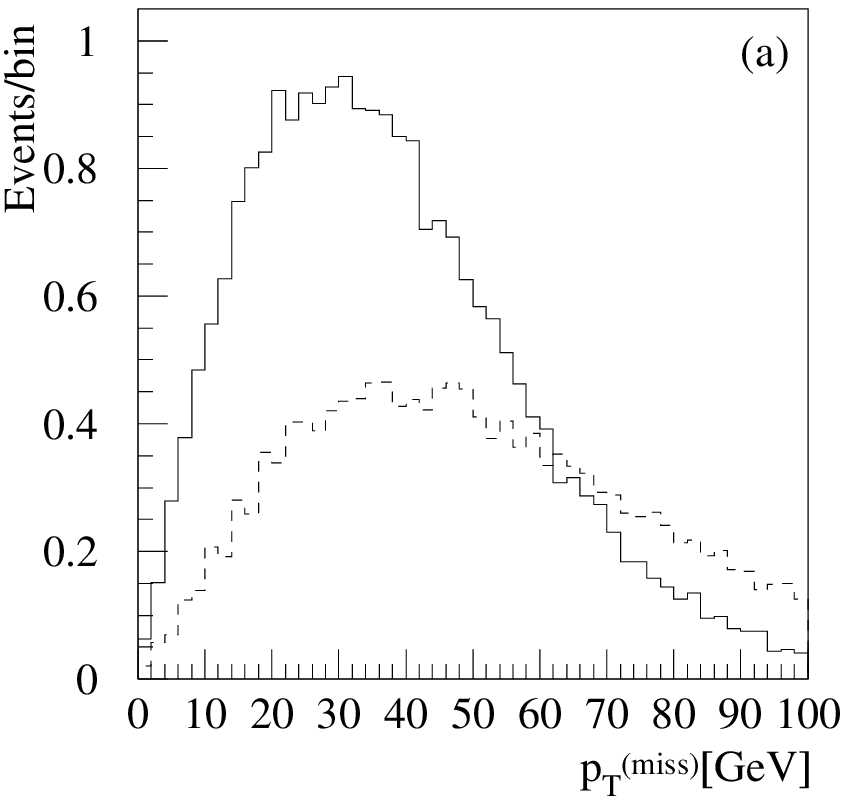,width=4.2cm}
\psfig{file=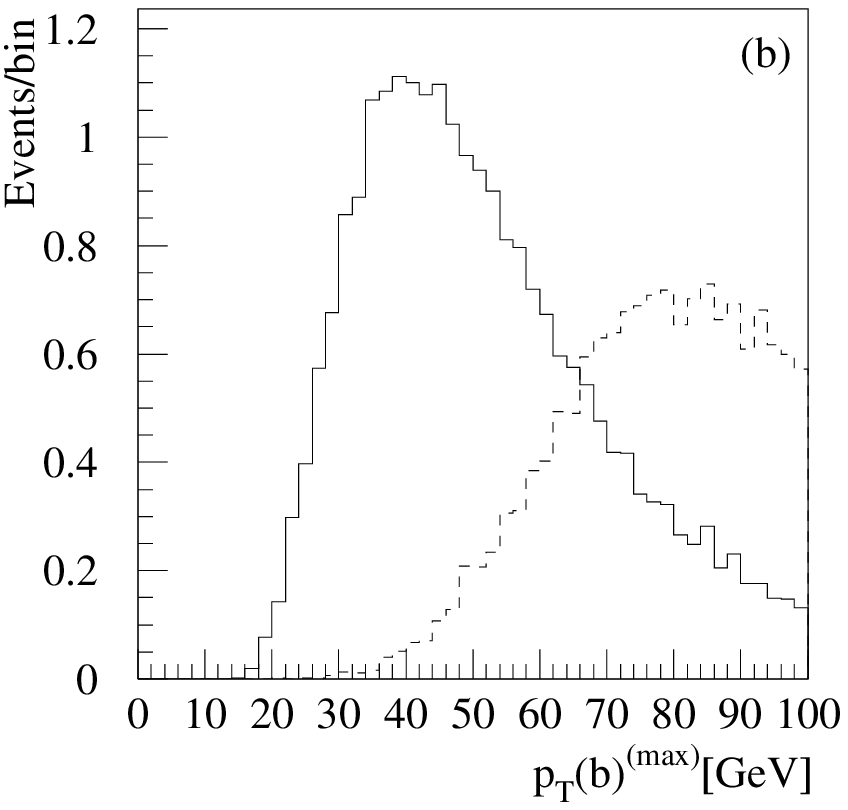,width=4.2cm}
}
\caption{\label{3b1l}Distributions
(a) of the missing transverse energy $\overlay{/}{p}_T$ and
(b) of the transverse momenta of the hardest $b$ jet in events with at least 
3 $b$ jets and at least 1 isolated lepton ($\geq 3b,\, \geq 1l$ signature).}
\end{figure}

In Fig.~\ref{3b1l}, we show the distribution of the
missing transverse energy and the transverse
momentum of the hardest $b$ jet in events with at least three tagged $b$
jets and at least one isolated lepton.
The $p_T$-distribution of the hardest $b$ shows 
that a more restrictive $p_T$ cut would
 bring no improvement. An upper cut on $p_T(b)$
enhances the signal-to-background ratio from
$S/B=23/15$ without cut to $S/B=19/3$ with $p_T(b)<70$~GeV for 
$2~\mathrm{fb}^{-1}$.
This cut also slightly improves the minimum
required luminosity for detection of the signal.

\begin{figure}[b]
\mbox{
\psfig{file=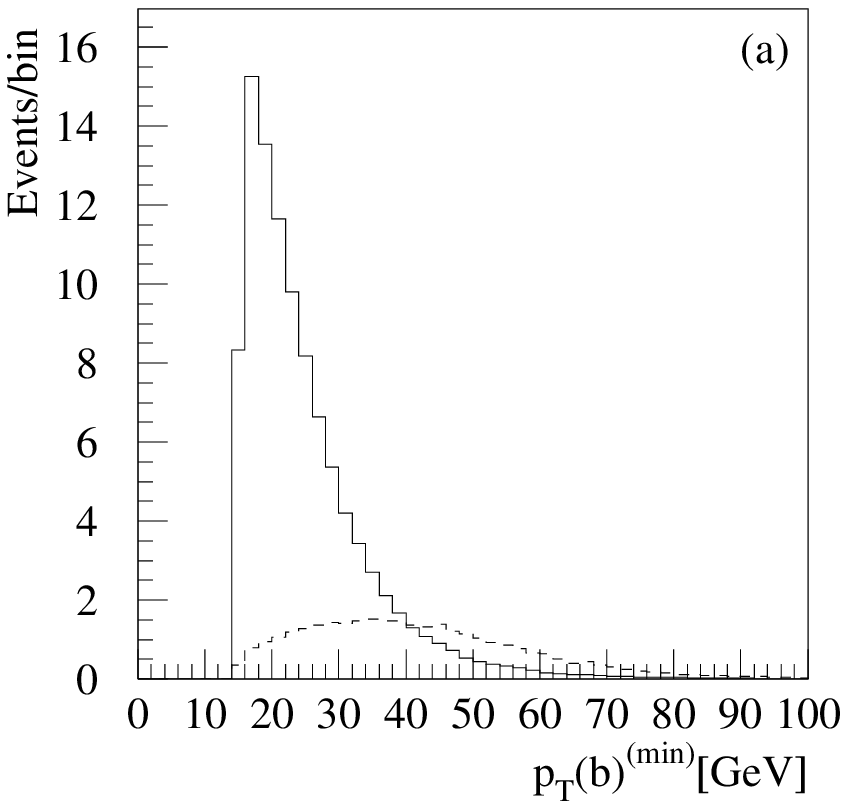,width=4.2cm}
\psfig{file=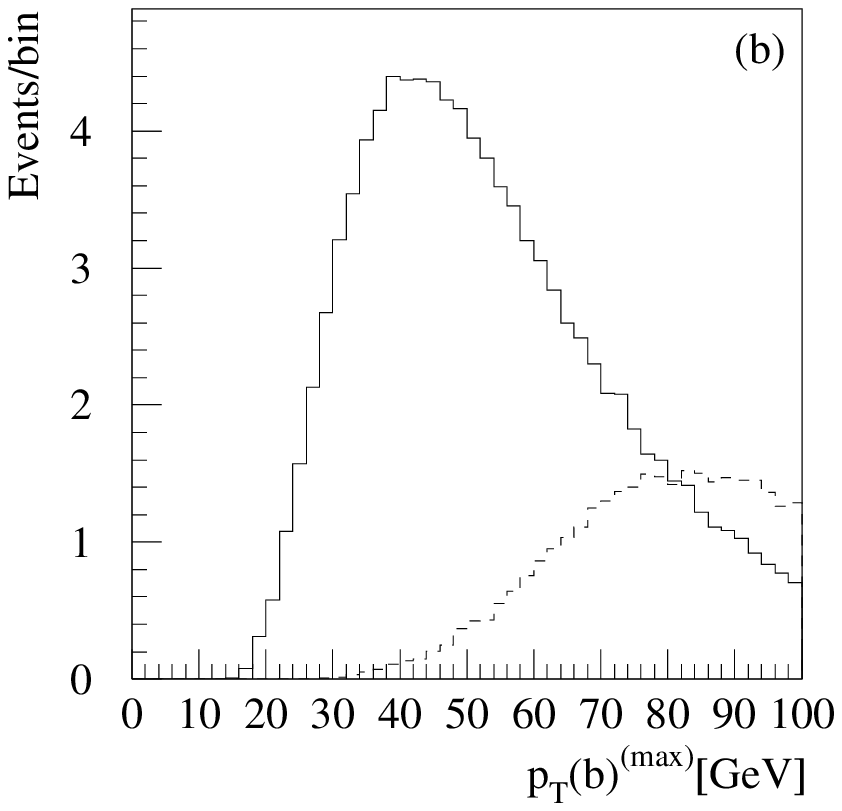,width=4.2cm}
}
\caption{\label{3bpt}Distributions of the transverse momenta (a) of the
third hardest $b$ jet and (b) of the hardest jet in events with at least 3 $b$
jets and $\overlay{/}{p}_T > 20$~GeV ($\geq 3b,\,\overlay{/}{p}_T$
signature).}
\end{figure}

Figure~\ref{3bpt} shows the transverse momenta of the third hardest
tagged $b$ jet and of the hardest jet in events of the
$\geq 3b,\,\overlay{/}{p}_T$ signature. The distribution of the third hardest $b$
in the SUSY signal is peaked at low $p_T$, so a lower
cut on the transverse momentum of the $b$ jets other than the basic cut
$p_T >15~\mbox{GeV}$ is not useful for this signature, too. 
But the transverse momentum
of the hardest $b$ of the background is significantly higher than in
the SUSY signal. Hence an upper cut on $p_T$ can improve the
signal-to-background ratio; for $ 2~\mathrm{fb}^{-1}$, $S/B = 100/31$
without a cut and $S/B = 76/5$ with an additional cut, $p_T(b)<70$~GeV.

\noindent
\underline{Discussion and conclusion}

\begin{figure}[t]
\psfig{file=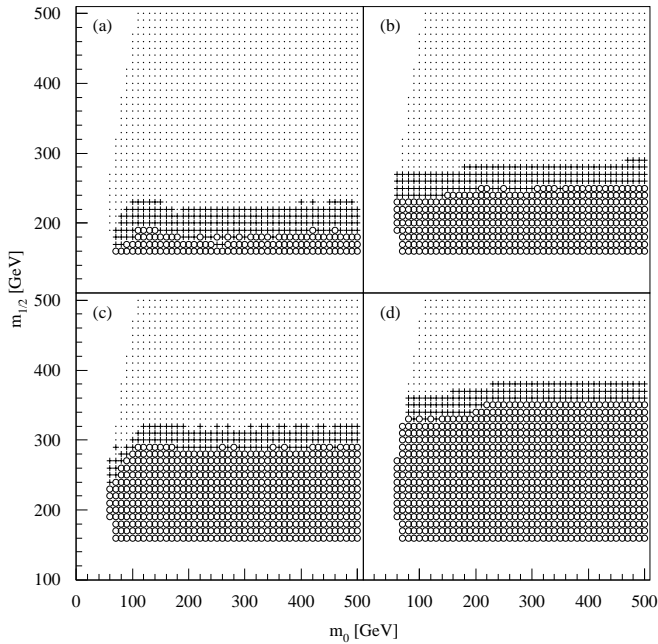,width=8.75cm}
\caption{\label{m0m12}Regions in the $m_0$-$m_{1/2}$ parameter space
where the signatures give a $3\sigma$ (pluses) and $5\sigma$
(circles) signal: (a) ($\geq 3b$, $\geq 1l$)
with $ 2~\mathrm{fb}^{-1}$, (b) ($\geq 3b$, $\overlay{/}{p}_T>20$~GeV)
with $ 2~\mathrm{fb}^{-1}$,
(c) ($\geq 3b$, $\geq 1l$) with 
$30~\mathrm{fb}^{-1}$ and (d) ($\geq 3b$, $\overlay{/}{p}_T>20$~GeV)
with $30~\mathrm{fb}^{-1}$. The other parameters are fixed at
$A_0=0$, $\tan\beta=10$ and $\mu>0$. The dots show mSUGRA points
that satisfy $m_{\tilde{\chi}^\pm_1}\,,m_{\tilde{e}_R} >100$~GeV,
$m_{\tilde{\tau}_1}>85$~GeV and $m_h>90$~GeV.}
\end{figure}

The regions in the $m_0$-$m_{1/2}$ parameter space where the
two signatures yield $3\sigma$ and $5\sigma$ evidence are
shown in Fig.~\ref{m0m12}.
The signals depend mainly on $m_{1/2}$ and only weakly on
$m_0$ because $m_{1/2}$ determines the masses of the 
neutralinos and charginos produced (see e.g. Ref.~\cite{kao}).

For large $m_0 \gtrsim 100$~GeV
the $3\sigma$ ($5\sigma$) reach of Tevatron Run II with
$ 2~\mathrm{fb}^{-1}$ is $m_{1/2}\approx 230$~GeV (190~GeV)
for the signature with at least 3 $b$ jets and at least one isolated lepton.
A luminosity of $30~\mathrm{fb}^{-1}$ can improve this reach 
to $m_{1/2}\approx 320$~GeV (290~GeV).
In the signature with at least 3 $b$ jets and $\overlay{/}{p}_T>20$~GeV, 
the signal is potentially larger. The $3\sigma$ ($5\sigma$) reach
 is $m_{1/2}\approx 290$~GeV (250~GeV) 
for $ 2~\mathrm{fb}^{-1}$
which improves to 380~GeV (350~GeV) for $ 30~\mathrm{fb}^{-1}$.

In summary, we have demonstrated that 
neutrino mass generation via $R$-parity violating SUSY can be tested 
at the Tevatron collider. 
The broken $R$-parity interactions are a simple and natural way 
of generating neutrino masses and provide a theoretical
 alternative to the see-saw mechanism. 
The neutrino masses and mixings indicated by 
the experimental results on neutrino 
oscillations can be explained with
$\lambda'_{133} \ll \lambda'_{i33} \sim 10^{-4}$, $i=2,3$.     
Because of the smallness of these couplings, 
the production and decay of sparticles proceeds via
the $R$-parity conserving channels except for the
decay of the LSP which decays into a $b\bar{b}$ pair and a neutrino in
most of the SUSY parameter space. 
At the Tevatron collider, the clean 
signature of at least three tagged $b$ jets and
at least one isolated lepton can be probed at $3 \sigma$ 
up to $m_{1/2} \approx 230$~GeV (320~GeV) with 
$2~\mathrm{fb}^{-1}$ ($30~\mathrm{fb}^{-1}$).
Another possible signature with at least three tagged $b$ jets and 
missing energy $\overlay{/}{p}_T>20$~GeV may improve the reach
up to $m_{1/2} \approx 290$~GeV (380~GeV) with $2~\mathrm{fb}^{-1}$ ($30~\mathrm{fb}^{-1}$) at $3 \sigma$. The physics reach in these channels 
is summarized in Fig.~\ref{m0m12}.

{\it Acknowledgments}:
We thank S. Dasu and S. Pakvasa for discussions. 
This research was supported by the U.S.~DOE
under Grant No.~DE-FG02-95ER40896 
and by the WARF. 
S.H. is supported by DFG under contract No.~HE~3241/1-1.

\end{document}